\documentstyle[preprint,eclepsf]{jpsj}

\def\runtitle{Effects of the Backward Scattering 
in Two-Dimensional Electron System}
\def\runauthor{Masakazu {\sc Murakami} and Hidetoshi {\sc Fukuyama}}

\title
{Effects of the Backward Scattering \\
in Two-Dimensional Electron System}
\author
{ 
Masakazu {\sc Murakami}\footnote{E-mail: murakami@watson.phys.s.u-tokyo.ac.jp}
and Hidetoshi {\sc Fukuyama}
}

\inst
{
Department of Physics, Tokyo University, Tokyo 113
}

\recdate
{
October 14, 1997
}

\abst
{Effects of the backward scattering with large momentum transfer are
examined in two-dimensional electron system with a special emphasis
on electrons around ($\pi$,0), (0,$\pi$).
The phase diagram is shown 
in the plane of temperature $T$
and hole doping $\delta$ in the mean field approximation and 
it is found that $d$-wave superconductivity,  
antiferromagnetism and $\pi$-triplet pair can coexist near half filling. 
}

\kword
{two-dimensional system, very flat band, backward
scattering, $d$-wave superconductivity, antiferromagnetism, $\pi$-triplet pair
}

\begin{document}
\sloppy
\maketitle

In recent years the electronic states in two-dimensional systems 
in the hole-doped copper oxide 
high-$T_{{\rm c}}$ superconductors have been studied intensively.
Especially, the anomalies in the normal state around ($\pi$,0) and (0,$\pi$)
have been elucidated in the angle resolved photoemission spectroscopy (ARPES)
experiments.\cite{ARPES1}
Near optimal hole doping, a very flat dispersion of quasiparticle excitations 
near the Fermi level around ($\pi$,0) and (0,$\pi$) 
has been reported in materials such as YBa$_{2}$Cu$_{3}$O$_{7-\delta}$ (YBCO) 
or Bi$_{2}$Sr$_{2}$CaCu$_{2}$O$_{8+\delta}$ 
(BSCCO).\cite{ARPES2}
This flatness of dispersion
is more prominent than that obtained in the single-particle
band calculations such as local-density approximation (LDA).\cite{LDA1,LDA2}
This 'extended' saddle-point behavior has been attributed to the many-body
correlation effects based on the results 
of quantum Monte Calro (QMC) simulations,
\cite{Bulut}
and propagator-renormalized fluctuation-exchange (FLEX) approximation,
\cite{ca}
or to the dimple of the CuO$_{2}$ planes from the recent 
LDA calculations.\cite{LDA3}
On the other hand, in the underdoped region, 
a pseudogap of the $d_{x^{2}-y^{2}}$ symmetry 
has been observed around ($\pi$,0) and (0,$\pi$) 
above the superconducting critical temperature.\cite{pgap1,pgap2,pgap3}
It has been indicated that there is strong coupling between 
quasiparticle excitations near the flat band and 
collective excitations centered near ($\pi$,$\pi$).\cite{pgap4}

Theoretically, such experimental features 
that the Fermi level approaches ($\pi$,0) and (0,$\pi$)
as hole doping rate is increased from half filling 
can be described by introducing 
not only nearest-neighbor hopping $t$ 
but also next-nearest-neighbor hopping $t'$.
Actually studies
looking for the instabilities
with a special emphasis on ($\pi$,0) and (0,$\pi$)
in the 2D Hubbard model 
have been carried out, by use of the renormalization group method
\cite{Schulz,ttu} 
and QMC.\cite{Husslein} 
These studies have shown that $d$-wave superconducticity prevails 
over antiferromagnetism by the effect of $t'$.
In this letter, we consider the situation
that the Fermi level approaches ($\pi$,0) and (0,$\pi$)
as hole doping rate is increased from half filling.
With a special emphasis on the effects of the backward scattering
processes with large momentum transfer between two electrons near ($\pi$,0) 
and (0,$\pi$),
we search for possible ordered states in the mean field approximation.
It is found that three order parameters,
$d$-wave Cooper pair,
Neel order and $\pi$-triplet pair are stabilized, and 
the phase diagram in the plane of temperature, $T$, and hole doping rate, 
$\delta$, has been determined.
We take unit of $\hbar=k_{{\rm B}}=1$.

\begin{figure}
\epsfigure{file=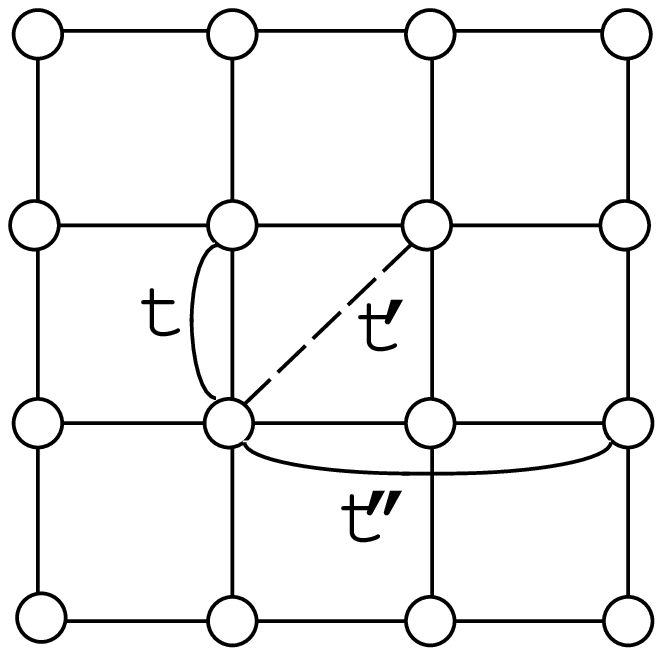 ,height=3cm}
\caption{Transfer integrals on a 2D square lattice;
$t$ (nearest neighbor), $t'$ (next nearest neighbor) and $t''$ 
(third neighbor).}
\label{hopping}
\end{figure}

We consider two-dimensional square lattice with the kinetic energy
given by,
\begin{eqnarray}
H_{0}&=& -\sum_{<ij>\sigma}t_{ij}\{c^{\dagger}_{i\sigma}c_{j\sigma}
+(\mbox{h.c.})\} - \mu\sum_{i}c^{\dagger}_{i\sigma}
c_{i\sigma},
\nonumber \\
&=& \sum_{p\sigma}\xi_{p}c^{\dagger}_{p\sigma}c_{p\sigma},
\label{eqn:h0}
\end{eqnarray}
where $t_{ij}$ is the transfer integral, $c_{i\sigma}(c^{\dagger}_{i\sigma})$
is the annihilation (creation) operator for the electron on the i-th site
with spin $\sigma$, and $\mu$ is the chemical potential.
The energy dispersion is given by
\begin{equation}
\xi_{p}=-2t(\cos p_{x}+\cos p_{y})-4t'\cos p_{x}\cos p_{y}-2t''(\cos 2p_{x}
+\cos 2p_{y})-\mu,
\end{equation}
including $t$ (nearest neighbor), $t'$ (next nearest neighbor) and $t''$ 
(third neighbor), as shown in Fig.~\ref{hopping}.

\begin{figure}
\epsfigure{file=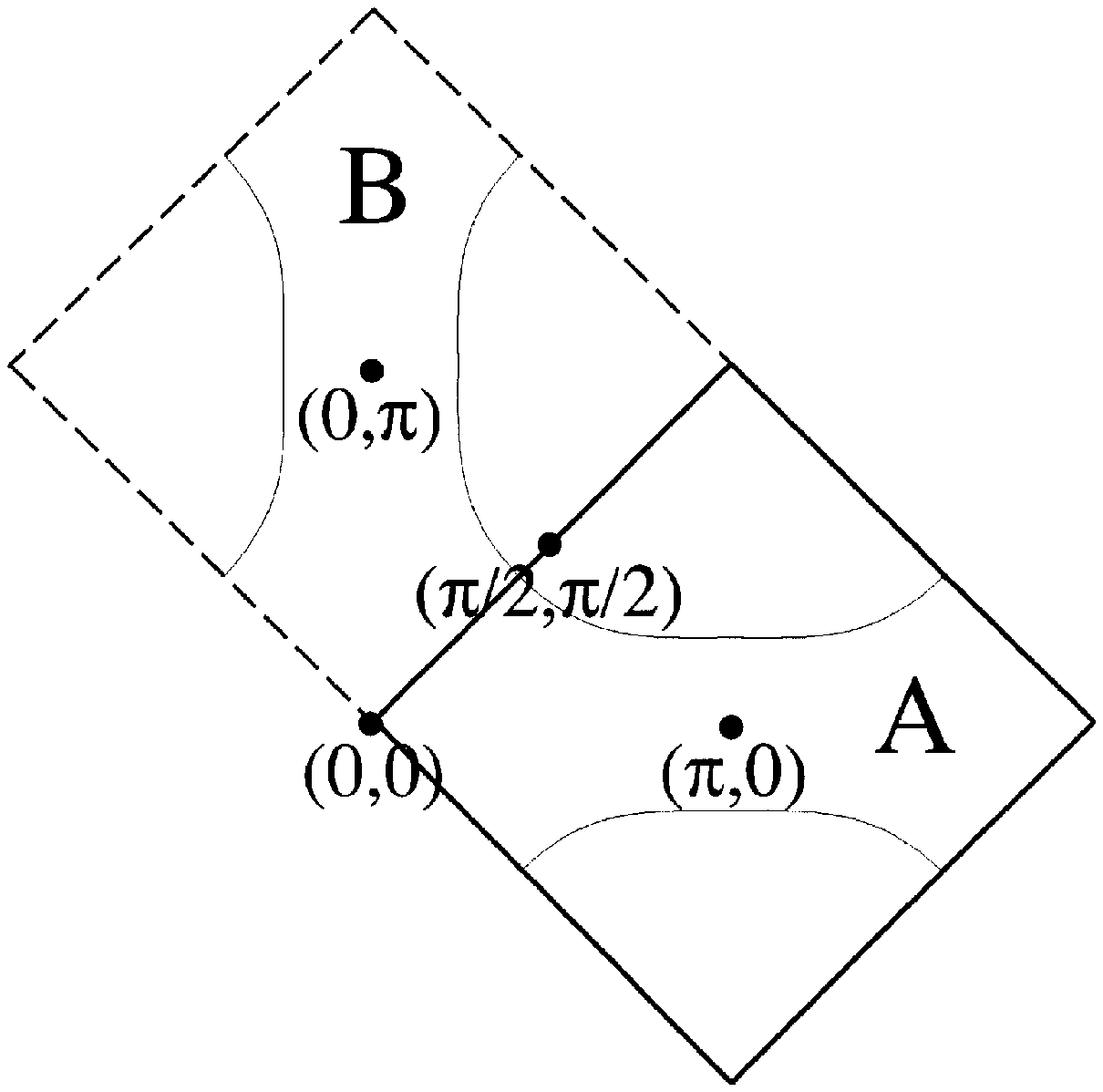 ,height=5cm}
\caption{The choice of the 1st Brillouin zone which consists of
the region A and B including
($\pi$,0) and (0,$\pi$), respectively, and
the Fermi surface in the half-filled case
for $t'/t=-1/6$ and $t''/t=1/5$.}
\label{2dbz}
\end{figure}

We take $t$ as the energy unit, i.e., $t=1$. The lattice constant is also 
taken as unity. We choose the values of $t'$ and $t''$ so that
the Fermi surface of YBCO type is reproduced near half filling.
In this letter, we take the following choice as
\begin{equation}
t'/t=-1/6,t''/t=1/5.
\end{equation}
In order to treat the scattering processes between two electrons
around ($\pi$,0) and (0,$\pi$) unambiguously,
we take the Brillouin zone 
as shown in Fig.~\ref{2dbz}, which consists of the region A and B 
including ($\pi$,0) and (0,$\pi$), respectively.
We note that $w_{p}\equiv \mbox{sgn}(\cos p_{y}-\cos p_{x})=+1(-1)$
for $p\in$A(B).
The Fermi surface in the half-filled case 
is also shown in Fig.~\ref{2dbz}.

\begin{figure}
\epsfigure{file=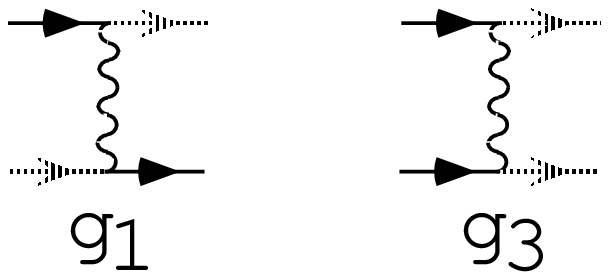 ,height=2cm}
\caption{Backward scattering processes with large momentum transfer.
The solid and dashed lines stand for electrons in the region 
A and B, 
respectively.}
\label{g1g3}
\end{figure}

Next we consider the interaction term.
We can introduce effective interaction between two electrons 
in the region A and B, just as in the g-ology 
in one-dimensional electron system.
Here we treat only the backward scattering with large momentum 
transfer, shown in Fig.~\ref{g1g3}, i.e., $g_{1}$ and $g_{3}$,
in analogy with normal and Umklapp processes in our previous work for
one-dimensional electron system
.\cite{ore}
We take only interaction between two electrons with antiparallel spins
into account, as in the Hubbard model, i.e., $g_{1}\equiv g_{1\perp}$ and
$g_{3}\equiv g_{3\perp}$.

\begin{figure}
\epsfigure{file=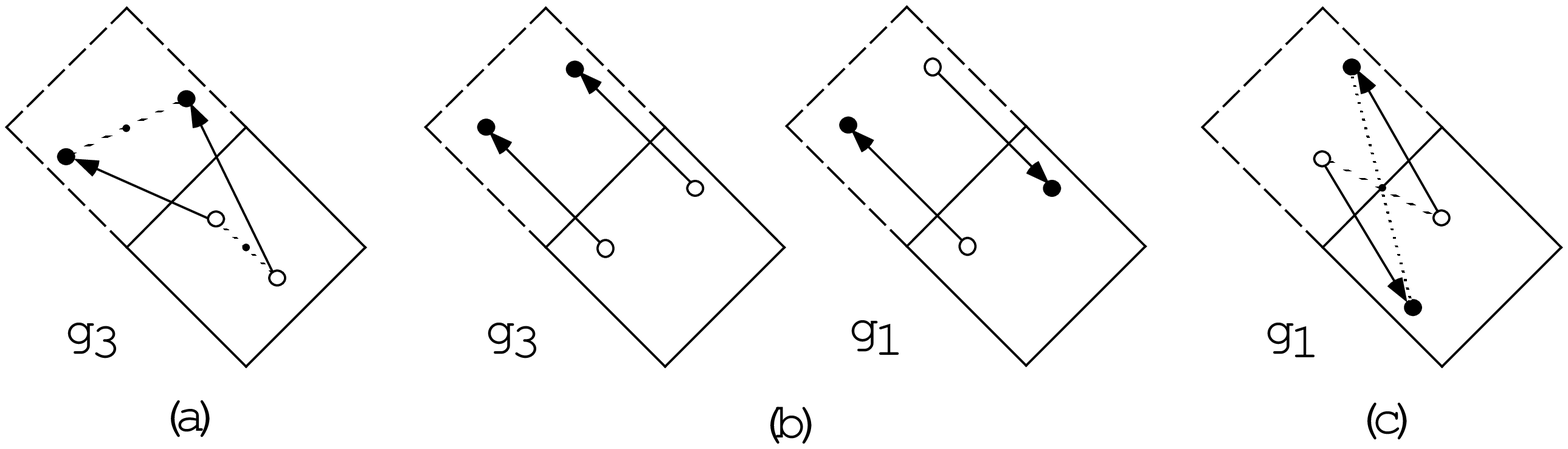 ,height=5cm}
\caption{Interaction processes considered in the mean field approximation. 
(a) Cooper channel, (b) density-wave channel, 
(c) $\pi$-pair channel.}
\label{process}
\end{figure}

We treat the above Hamiltonian in the mean field approximation under the
assumption $g_{1}=g_{3}\equiv g>0$ for simplicity.
By noting that there exist the scattering processes as shown 
in Fig.~\ref{process},
we naturally introduce the following order parameters, i.e.,
Cooper-pair with total momentum equal to zero 
in the original 1st Brillouin zone which can be expressed as follows,
\begin{subequations}
\begin{equation}
\Delta_{A}=g\mathop{{\sum}'}_{k}<c_{Q_{A}-k\downarrow}
c_{Q_{A}+k\uparrow}>,
\Delta_{B}=g\mathop{{\sum}'}_{k}<c_{Q_{B}-k\downarrow}
c_{Q_{B}+k\uparrow}>,
\end{equation}
and staggered carrier density of each spin,
\begin{equation}
\Delta_{\uparrow}=g\mathop{{\sum}'}_{k}<c_{Q_{B}+k\uparrow}^{\dagger}
c_{Q_{A}+k\uparrow}>,
\Delta_{\downarrow}=g\mathop{{\sum}'}_{k}<c_{Q_{B}-k\downarrow}
^{\dagger}
c_{Q_{A}-k\downarrow}>,
\end{equation}
and $\pi$-pair
\begin{equation}
\Delta_{+}=g\mathop{{\sum}'}_{k}<c_{Q_{A}-k\downarrow}
c_{Q_{B}+k\uparrow}>,
\Delta_{-}=g\mathop{{\sum}'}_{k}<c_{Q_{B}-k\downarrow}
c_{Q_{A}+k\uparrow}>,
\end{equation}
\end{subequations}
where $Q_{A}=(\pi,0)$ and $Q_{B}=(0,\pi)$, respectively, and
\[
\mathop{{\sum}'}_{k} \equiv \sum_{|k_{x}|+|k_{y}|<\pi}.
\]
It is to be noted that $Q_{A}\pm k$ and $Q_{B}\pm k$ always lie in the
region A and B, respectively, for $k$ satisfying $|k_{x}|+|k_{y}|<\pi$.
This is the reason for the choice of the Brillouin zone as shown 
in Fig.~\ref{2dbz}.
We take
\begin{subequations}
\begin{equation}
\Delta_{1} \equiv \Delta_{A}= -\Delta_{B},
\end{equation}
\begin{equation}
\Delta_{2}\equiv 2\Delta_{\uparrow} =-2\Delta^{\ast}_{\downarrow},
\end{equation}
\begin{equation}
\Delta_{3}\equiv \Delta_{+}=-\Delta_{-},
\end{equation}
\end{subequations}
so that each of the above three
order parameters is possible independently of each other
for $g>0$. 
They stand for $d$-wave Cooper-pair, Neel order and $\pi$-triplet pair,
respectively. We note that $\Delta_{2}$ is
real for $g_{1}=g_{3}$.
Our $g_{3}$ processes included in the Cooper channel are the same as
the 'pair-tunneling' of two electrons located
around ($\pi$,0) and (0,$\pi$), which favors
$d_{x^{2}-y^{2}}$ pairing, in the 2D Hubbard model.\cite{kuroki}
These interaction processes are not present as important factors
in the recent model by Assaad {\em et al.}, who
have introduced the additional 
interaction expressed as the square of the
single-particle nearest-neighbor hopping.\cite{imada}

We obtain the mean field Hamiltonian as follows:
\begin{subequations}
\begin{equation}
H^{MF}=\mathop{{\sum}'}_{k}
\psi^{\dagger}_{k}M_{k}\psi_{k}+H_{c},
\end{equation}
\begin{equation}
\psi^{\dagger}_{k}=(c^{\dagger}_{Q_{A}+k\uparrow},\:
c_{Q_{A}-k\downarrow},\:
c^{\dagger}_{Q_{B}+k\uparrow},\:c_{Q_{B}-k\downarrow}),
\end{equation}
\begin{equation}
M_{k} = \left(
\begin{array}{cccc}
a_{k} & -\Delta_{1} & -\Delta_{2} & \Delta_{3}\\
-\Delta^{\ast}_{1} & -a_{k} & -\Delta^{\ast}_{3} & -\Delta_{2}\\
-\Delta_{2} & -\Delta_{3} & b_{k} & \Delta_{1}\\
\Delta^{\ast}_{3} & -\Delta_{2} & \Delta^{\ast}_{1}  & -b_{k}
\end{array}
\right),
\end{equation}
\begin{equation}
a_{k}\equiv \xi_{Q_{A}\pm k},b_{k}\equiv \xi_{Q_{B}\pm k},
\end{equation}
\begin{equation}
H_{c}=\frac{1}{g}\{2(|\Delta_{1}|^{2}+|\Delta_{3}|^{2})+(\Delta_{2})^{2}\}-\mu.
\end{equation}
\end{subequations}
There are four quasiparticle energy bands, $\pm E_{+}(k)$ and $\pm E_{-}(k)$,
\begin{subequations}
\begin{equation}	
E_{\pm}(k)=\sqrt{\frac{a_{k}^{2}+b_{k}^{2}}{2}+
|\Delta_{1}|^{2}+(\Delta_{2})^{2}+|\Delta_{3}|^{2}
\pm A(k)},
\end{equation}
\begin{full}
\begin{equation}	
A(k)\equiv\sqrt{(a_{k}-b_{k})^{2}\left [ \left ( \frac{a_{k}+b_{k}}{2} \right 
)^{2}+
|\Delta_{3}|^{2} \right ]+[(a_{k}+b_{k})(\Delta_{2})-2\mbox{Re}(\Delta_{1}^{*}
\Delta_{3})]^{2}}.
\end{equation}
\end{full}
\end{subequations}
The self-consistent equations are given by
\begin{full}
\begin{subeqnarray}
\Delta_{1}&=&
\frac{g}{4}\mathop{{\sum}'}_{k}\sum_{\alpha = \pm}
\frac{1}{E_{\alpha}(k)}\tanh\frac{E_{\alpha}(k)}{2T} \left\{
\Delta_{1}+\alpha\frac{2\Delta_{3}}{A(k)}\left[\mbox{Re}(\Delta_{1}^{*}
\Delta_{3})-a_{k}\Delta_{2} \right ]\right\},
\\
\Delta_{2}&=&\frac{g}{2}\mathop{{\sum}'}_{k}\sum_{\alpha = \pm}
\frac{1}{E_{\alpha}(k)}\tanh\frac{E_{\alpha}(k)}{2T} \left\{
\Delta_{2}\left [1+\alpha\frac{(a_{k}+b_{k})^{2}}{2A(k)}\right ]
-\alpha\frac{2\mbox{Re}(\Delta_{1}^{*}
\Delta_{3})a_{k}}{A(k)}\right\},
\\
\Delta_{3}&=&\frac{g}{4}\mathop{{\sum}'}_{k}\sum_{\alpha = \pm}
\frac{1}{E_{\alpha}(k)}\tanh\frac{E_{\alpha}(k)}{2T} \left\{
\Delta_{3}\left [1+\alpha\frac{(a_{k}-b_{k})^{2}}{2A(k)}\right ]
+\alpha\frac{2\Delta_{1}}{A(k)}\left[\mbox{Re}(\Delta_{1}^{*}
\Delta_{3})-a_{k}\Delta_{2} \right ]\right\},
\\
n&=&1-\frac{1}{2}\mathop{{\sum}'}_{k}\sum_{\alpha = \pm}
\frac{1}{E_{\alpha}(k)}\tanh\frac{E_{\alpha}(k)}{2T}\left\{
(a_{k}+b_{k})\left\{1+\frac{2\alpha}{A(k)}\left[
(\frac{a_{k}-b_{k}}{2})^{2}+\Delta_{2}^{2}\right]\right\}
-\alpha\frac{4\Delta_{2}\mbox{Re}(\Delta_{1}^{*}
\Delta_{3})}{A(k)} \right\},
\end{subeqnarray}
\end{full}
where $n$ is the electron filling, which is related to hole doping $\delta$
by $\delta = 1-n$.

\begin{figure}
\epsfigure{file=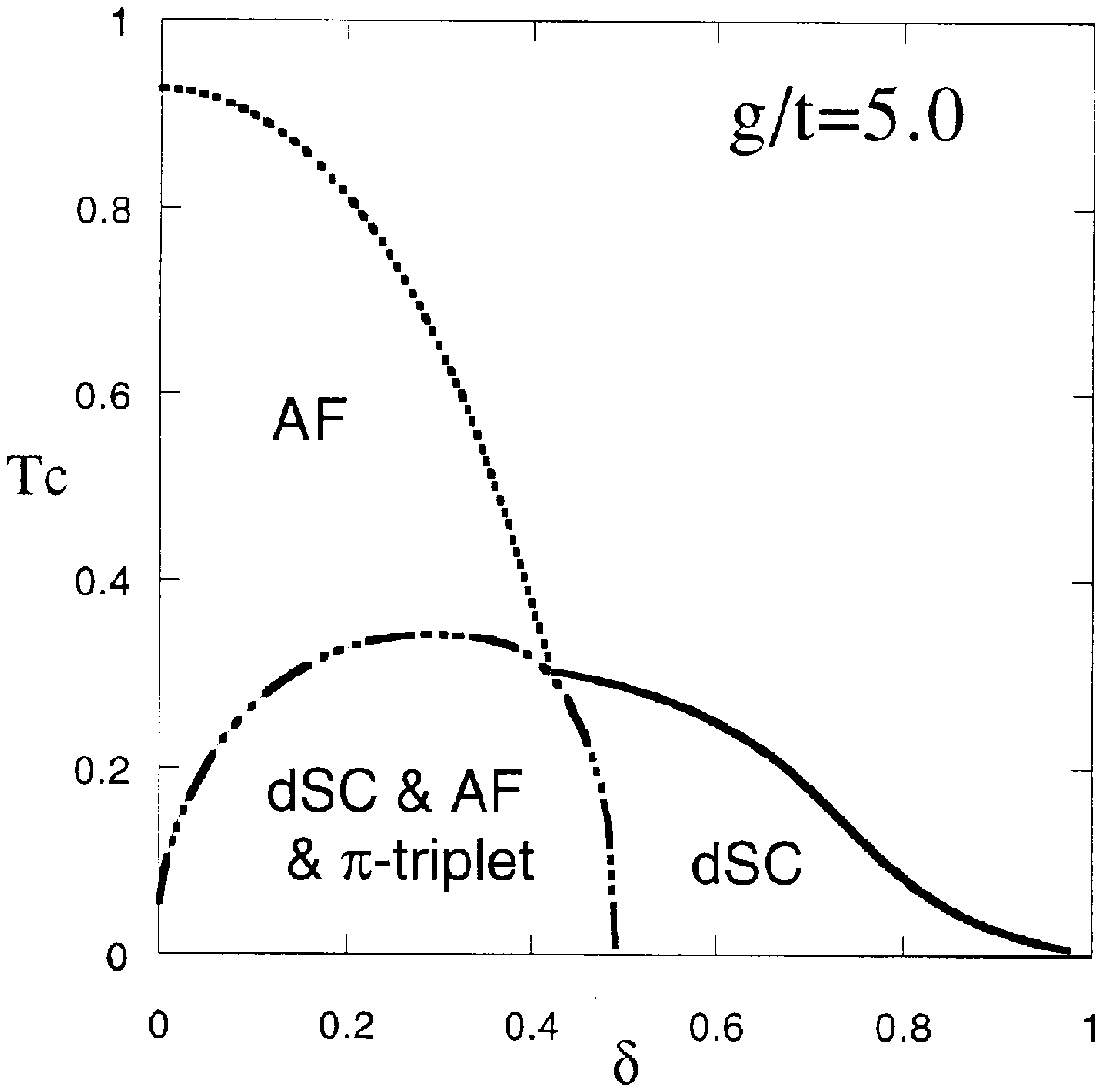 ,height=6cm}
\caption{The phase diagram in the plane of $T$ and $\delta$ for 
a choice of $g/t=5.0$.}
\label{phasediagram}
\end{figure}

These equations are solved numerically, under the assumption that
both $\Delta_{1}$ and $\Delta_{3}$
are real.
The resultant phase diagram is shown in Fig.~\ref{phasediagram}
in the plane of $T$ and $\delta$ for a choice of $g/t=5.0$.
In this case we find that $d$-wave superconductivity (dSC), 
antiferromagnetism (AF) and 
$\pi$-triplet pair can coexist near half filling.
Such a close relationship between dSC and AF has been indicated by
SO(5) theory.\cite{SCZhang}
We note that if two of three order parameters coexist, 
another one always results.
It is noted that the stability of
antiferromagnetism near half filling is not due to the nesting but due to
scattering processes involving both $g_{1}$ and $g_{3}$.
 
In summary, we have studied possible ordered states 
of interacting electrons on a square lattice
with a special emphasis
on the backward scattering, $g_{1}$ and $g_{3}$, i.e.,
'exchange' and 'Umklapp' processes, respectively, between two electrons around
($\pi$,0) and (0,$\pi$) for the case of YBCO type Fermi surface.
If we take $g_{1}=g_{3}>0$, $d$-wave superconductivity (dSC), 
antiferromagnetism (AF) and 
$\pi$-triplet pair can coexist near half filling.
On the other hand, the results of our preliminary calculations
indicate that for $g_{1}\neq g_{3}$ 
dSC can compete with AF and especially for $g_{1}=0$ 
dSC prevails over AF in the half-filled case.
On the other hand, in the case of electron-doped materials such as 
Nd$_{2-x}$Ce$_{x}$CuO$_{4+\delta}$ (NCCO), 
the Fermi level lies far above ($\pi$,0)
and (0,$\pi$), but gets close to ($\pm \pi /2$, $\pm \pi /2$),\cite{ARPES2}
for which recent QMC calculations show that 
not only $d_{x^{2}-y^{2}}$ but also $d_{xy}$ pairing correlation 
are enhanced in the 2D Hubbard model.\cite{kuroki}
These results indicate that there exists a variety depending on
the shape of the Fermi surface and scattering processes,
which will be reported in detail elsewhere.

M.M. would like to express his gratitude to Hiroshi Kohno and Kazuhiko
Kuroki for instructive 
discussions and suggestions.  
This work is financially supported by a Grant-in-Aid
for Scientific Research on Priority Area "Anomalous Metallic State near the 
Mott Transition" (07237102) from the Ministry of Education, Science, Sports
and Culture.

\end{document}